\documentclass{scrartcl}
\usepackage{amsmath}
\usepackage{graphicx}

\title{Common origin of superluminal neutrinos and DAMA annual modulation}
\author{ Gerald Xavier Gilbert-Thorple$^*$, J\^os\'e J.\ Jesus$^\dagger$\\ 
  \textit{$^*$Kavli University of Southern North Dakota,
Hoople, ND, USA}\\
  \textit{$^\dagger$Beijing University, Athens Campus, 
Athens, Greece, PRC}
}
\date{}

\begin{document}
\maketitle

\begin{abstract}
We suggest that the recently observed superluminal neutrino
propagation observed by OPERA may have an environmental origin
that can simultaneously explain the DAMA/LIBRA annual modulation.
Our proposal can be unambiguously tested by the prediction that
OPERA will also observe an annual modulation in the speed of
neutrinos.
Experiments geographically distant from Gran Sasso will however not be
able to observe these effects.  We do not rely upon any violation of
Lorentz symmetry except for spontaneous breaking of 
translational invariance.  We discuss implications for the CRESST
experiment, also located at Gran Sasso.

\end{abstract}

In a {\it tour de maitre} of striking disproportions, the physics world
and indeed fragments of the general public have been astonished by the 
long-awaited overthrow of relativistic despotism implied by the OPERA
experiment's recent detection of muon neutrinos traveling faster than
light between CERN and Gran Sasso National Laboratory (LNGS)
\cite{:2011zb}.   Before declaring such a victory
however, we should be careful to rule out more mundane explanations
such as the one proposed in the present paper (see below).  It is
intriguing that at the same time (plus or minus ten years) the
DAMA/LIBRA experiment, seemingly coincidentally located at 
Gran Sasso,
has been reporting equally convincing evidence for the annual
modulation of unidentifiable backgrounds in their detector
\cite{Bernabei:2008yi}, which
could include the impotent interactions of dark matter particles blowing
through their apparatus at highly subluminal speeds.  

The dramatic discovery of OPERA has been met with a remarkable paucity
of plausible explanations or criticisms from the theory community 
\cite{AmelinoCamelia:2011dx}-\cite{Moffat:2011ue}.  In a remarkably
cunning bit of sophistry, Cohen {\it et al.} \cite{Cohen:2011hx}
argued that superluminal neutrinos would lose too much energy via
decay to be consistent with the observation, but this hypothesis is
clearly ruled out by the observations of ref.\ \cite{a:2011zb}. 
Similarly ref.\  \cite{Giudice:2011mm} has pointed out theoretical
difficulties for superluminal propagation, but their analysis is not
self-consistent since the authors clearly knew about the experimental
results before they were reported, showing evidence for superluminal
propagation of information.

Our observation is inspired by the curious proposal of ref.\ 
\cite{Dvali:2011mn}, which claimed that an extra graviton polarization
sourced by the earth could account for the superluminal motion. This
claim seems absurd since as is well known, passage through a
gravitational potential can only make particles travel slower, not
faster than light\footnote{We thank Dr.\ G.D.\ Morose for charming
diatribes on this subject}, unless the source has unusual properties.
Here we posit the existence of a source that violates the weak null
energy condition (WNEC),\footnote{We remind the reader that ``null''
in the context of WNEC modifies ``condition'' rather than ``energy''}
which is also a requirement for sourcing traversible wormhole
solutions in general relativity \cite{Morris:1988tu}.  A wormhole
with one end located in close proximity to LNGS would be consistent
with the explanation of ref.\ \cite{Dvali:2011mn}, while removing 
the objections to a power-law behavior for the energy-dependence
of the superluminal effect, since the experiments which would
seemingly contradict this hypothesis \cite{Kalbfleisch:1979rm}
were not located at LNGS.

Emanating from this wormhole will be a number of objects which do not
get very far inside the Gran Sasso Mountain, but in addition we expect a
contribution to the dark matter wind that can have a significant
impact on the annual modulation observed by DAMA.  This observation is
especially timely since it has been recently pointed out that there is
a strong correlation between the DAMA observations and the annually
modulated flux of atmospheric neutrinos \cite{Blum:2011jf}. It is
important to notice that the extra component to the dark matter wind
emanating from the wormhole will also be annually modulated by the
stretching of the wormhole over the course of a year (except in the
unlikely event that the other end of the wormhole lies along the  axis
of rotation of the earth's orbit).  We believe that this effect should
exactly compensate the atmospheric muon background not taken into
account by DAMA's analysis, meaning that their net modulation comes
from the ordinary (non-wormhole) component of the dark matter wind. 
Here we have some connection with the proposal of ref.\ 
\cite{Hannestad:2011bj}, which fantasized the propagation of sterile
 neutrinos mixing with $\nu_\mu$
in warped extra dimensions; our wormhole too could be traversing in
part the extra dimensions.  However unlike ref.\
\cite{Hannestad:2011bj} we do not need to assume that the mixing angle
of $\nu_\mu$ with the sterile neutrino is simultaneously both large and
small.

Unlike other proposed explanations of the OPERA anomaly, our proposal
is fully testable, since the annual distortions of the wormhole will
change the gravitational potential transmitted by the extra graviton
polarization, leading to annual modulations in the $\nu_\mu$ velocity
anomaly of one part in $10^{10}$, which could be measured by OPERA
given a sufficiently long period of data collection and analysis.  In addition,
there are implications for the CRESST experiment which is also located
at LNGS.  These and other details of the present analysis, complete
with equations, will be discussed in a forthcoming publication 
\cite{wip} now that we have laid claim to our idea.

\section*{Acknowledgment}

We thank members of the OPERA and DAMA collaborations for candid
discussions admitting the probability of our explanation.

\end{document}